\documentclass [12pt]{iopart}

\usepackage{amssymb}
\usepackage{iopams}
\usepackage{graphicx}
\usepackage{cite}

\newcommand{\vcm}{\mathbf{m}}
\newcommand{\const}{\mathrm{const}}

\begin{document}

\title[Multidimensional purity-bounded uncertainty
relations]{Purity-bounded uncertainty relations in
multidimensional space --- generalized purity}

\author{M Karelin}
 \address{B.~I.~Stepanov Institute of Physics, National Academy of
Sciences, Minsk, 220072 Belarus}
 \ead{karelin@dragon.bas-net.by}








\begin{abstract}
Uncertainty relations for mixed quantum states (precisely,
purity-boun\-ded position-momen\-tum relations, developed by
Bastiaans and then by Man'ko and Dodonov) are studied in general
multi-dimensional case. An expression for family of mixed states
at the lower bound of uncertainty relation is obtained. It is
shown, that in case of entropy-bounded uncertainty relations,
lower-bound state is thermal, and a transition from
one-dimensional problem to multi-dimensional one is trivial.
Results of numerical calculation of the relation lower bound for
different types of generalized purity are presented. Analytical
expressions for general purity-bounded relations for highly mixed
states are obtained. PACS number: 03.65.Ca.
\end{abstract}


\section{Introduction and review}
\label{Intro}

Well-known position-momentum uncertainty relation for standard
deviations of $\hat x$ and $\hat p$ operators,
 \begin{equation}
 \Delta x \Delta p \geqslant \hbar/2,
 \label{eq-triv}
 \end{equation}
is valid for any state (described either by a wavefunction or by a
density matrix \cite{Kholevo1982,StolerNewman1972}) and plays an
quite an important role in quantum physics. In particular,
uncertainty relation sets the precision limits of measurement
process for non-commuting observables \cite{Mensky2000,Resch2004}.
Another important example is that generalized coherent states (and
squeezed states) could be defined as a set of states which
minimize an uncertainty relation, see \cite{Trifonov1999}.
Uncertainty principle and properties of its minimum is also of
special interest in theory of operators in Hilbert space, see
further references in a recent work by Goh and
Miccelli~\cite{GohMicchelli2002}.

The inequality (\ref{eq-triv}) have been generalized to include
extra dependence on degree of purity \cite{Leonardt1997} of a
quantum state
 \begin{equation}
 \mu = \Tr(\hat\rho^2)
 \label{eq-mu2}
 \end{equation}
($\hat\rho$ is a density operator), 
the parameter $0 \leqslant \mu \leqslant 1$ and equality $\mu = 1$
is achieved only for pure states. An asymptotic inequality for
one-dimensional highly mixed states with $\mu \ll 1$ has a form
\cite{Bastiaans1983a,Bastiaans1983b,Bastiaans1984,Bastiaans1986a,%
Bastiaans1986b,DodonovManko1989,Dodonov2002}
 \begin{equation}
 \Delta x \Delta p \geqslant \frac{8}{9 \, \mu} \, \frac{\hbar}{2}.
 \label{eq-1d}
 \end{equation}
In addition to the trace (\ref{eq-mu2}) of squared density
operator, there are other measures of overall purity (see above
cited papers for details, especially a recent comprehensive review
on purity-bounded relations \cite{Dodonov2002}).

Another approach for treatment of uncertainty relation for
mixed states are developed, by Wolf, Ponomarenko,
Agarwal~\cite{PonomarenkoWolf2002,AgarwalPonomarenko2003},
and also by Vourdas and his co-authors~\cite{Vourdas2004,%
ChountasisVourdas1998}. In the cited works, the uncertainty
relation is expressed in terms of correlations of respective
observables. On the other hand, the inequality of the type
(\ref{eq-1d}) relates uncertainties in conjugated variables and a
measure of overall purity of state.

A generalization of the uncertainty relation (\ref{eq-triv}) to
multidimensional space (vector observables, which can appear e.~g.
for multimode states, or multi-particle situations) was
investigated in early days of quantum mechanics~%
\cite{Robertson1934} (see also a review in
\cite{DodonovManko1989}) and is still drawing attention of
researches \cite{Trifonov2000,Sudarshan_etal1995,Adesso_etal2004}.
In its most simple form, uncertainty relation for $n$-dimensional
position and momentum operators $\hat X = (\hat x_1, \hat x_2,
\ldots, \hat x_n)$, $\hat P = (\hat p_1, \hat p_2, \ldots, \hat
p_n)$ could be written as
 \begin{equation}
 \label{eq-triv-nd}
 (\Delta X \Delta P)^n \geqslant \left(\frac{\hbar}{2}\right)^n,
 \end{equation}
with definitions
 \[
 (\Delta X)^2 = \frac{1}{n} \prod\limits_{i=1}^n \Delta x_i,
 \qquad
 (\Delta P)^2 = \frac{1}{n} \prod\limits_{i=1}^n \Delta p_i.
 \]
In fact, due to equality between different coordinates in minimum
of uncertainty relation, the inequality (\ref{eq-triv-nd}) has the
same meaning as $n$-th degree of standard one-dimensional relation
(\ref{eq-triv}), see also discussion in \cite{KarelinLazaruk1998}.


The problem of generalization of the inequality (\ref{eq-1d}) for
multidimensional case was treated in papers by Karelin and
Lazaruk~\cite{KarelinLazaruk1998,KarelinLazaruk2000}.
In particular, in our first paper on this topic
\cite{KarelinLazaruk1998}, it has been shown with the help of the
Wigner function formalism, that there is a non-trivial dependence of
the purity-bounded uncertainty relation limit on the number of
dimensions. For highly mixed states with $\mu \ll 1$,
 \begin{equation}
 (\Delta X \Delta P)^n \geq
 \frac{C(n)}{\mu} \left(\frac{\hbar}{2}\right)^n,
 \quad
 C(n) = \frac{2^{n+1} (n+1)!}{(n+2)^{n+1}},
 \label{eq-asymp}
 \end{equation}
where the parameter $C(n)$ characterizes distance from a minimum
$(\hbar/2)^n$ for pure states. In deriving (\ref{eq-asymp}), we
assumed, that the Wigner function of the minimum-uncertainty state
is nonnegative.


In another paper \cite{KarelinLazaruk2000}, the structure of the
density matrix near the lower bound of uncertainty relation was also
found, using decomposition of the density matrix in terms of Fock
states (which form an orthogonal basis with a minimal uncertainty)
 \begin{equation}
 \hat\rho = \sum_{\vcm,\vcm'}
    a_{\vcm,\vcm'} |m_1\rangle \, |m_2\rangle \ldots |m_n\rangle \:
         \langle m'_1| \, \langle m'_2| \ldots \langle m'_n|,
 \label{eq-Fock}
 \end{equation}
with the additional condition
 \[
 \displaystyle \mu =
 \sum\limits_{\vcm,\vcm'} {\left|a_{\vcm,\vcm'}\right|^2 } =
 \const,
 \]
where $\vcm=(m_1,\ldots,m_s)$ is a ``vector'' index with integer
nonnegative components.

At the lower bound of uncertainty relation, density matrix in Fock
representation is diagonal, $a_{\vcm,\vcm'}=a_{\vcm,\vcm}
\delta_{\vcm,\vcm'}$; coefficients $a_{\vcm,\vcm}$ depend linearly
on the `norm' of vector index $\left\|\vcm\right\| =
\sum_{i = 1}^n{m_i}$, and number of Fock states in representation of
$\hat\rho$ is finite. Coefficients $a_{\vcm,\vcm}$ of this
decomposition are degenerate, and their multiplicity is determined
by norm $\|\vcm\|$ and dimensionality $n$:
 \begin{equation}
 \label{eq-degener}
 g_{\|\vcm\|}^{(n)} =
 \frac {\left(\|\vcm\| + n - 1\right)!}
 {\left(\|\vcm\|\right)!\left(n - 1\right)!}.
 \end{equation}
The inequality obtained in \cite{KarelinLazaruk2000} correctly
describes the whole range of $\mu$, including perfectly pure
case $\mu = 1$. In particular, in the interpolating form
it becomes
 \begin{equation}
 \Delta X \Delta P \geqslant
     \frac{\hbar}{2} \, \frac{n+2\,L(\mu)}{n+2}
 \label{eq-intrp-nd}
 \end{equation}
with the auxiliary real parameter $L(\mu)$ being a root of
the transcendental equation
 \begin{equation}
 \mu = \frac{(n+2\,L)(n+1)! \,\Gamma(L)}{(n+2)\,\Gamma(L+n+1)},
 \label{eq-mu-condition}
 \end{equation}
where $\Gamma(y)$ is Euler's gamma-function.

It is also necessary to note that the inequality, mathematically
practically the same as uncertainty relation, but with another
physical meaning, is often used for classical wave fields, e.~g.
in optics \cite{Bastiaans1983a,Bastiaans1983b,Bastiaans1984,%
Bastiaans1986a,Bastiaans1986b,KarelinLazaruk1998}. Results of the
present article, as well as of preceding papers
\cite{KarelinLazaruk1998,KarelinLazaruk2000} could be used, with
appropriate change of notations, for classical partially coherent
fields and sources (in $1$-, $2$- and $3$-dimensional space
\cite{KarelinLazaruk2002}).

The uncertainty relations (\ref{eq-asymp}), (\ref{eq-intrp-nd})
could be further generalized in order to take into account the
dependence of inequality minimum on eigenvalues of density
operator. Preliminary report on this topic, with stress on
partially coherent classical fields, was published in
\cite{Karelin2002}. Obtaining such a relation, together with the
study of its asymptotics, is the main aim of the present paper.

 \section{Uncertainty relation for the diagonal representation
 of the density matrix}
 \label{unc-decomp}

Any density matrix $\hat\rho$ has a spectral decomposition
\cite{Feynmann1972}
 \begin{equation}
 \label{eq-spectral}
 \hat\rho=\sum_m \rho_m \, |\psi_m\rangle \, \langle\psi_m |,
 \end{equation}
where $\rho_m$ are the eigenvalues, and $|\psi_m\rangle$ are
eigenvectors of the density operator, then, each of vectors
$|\psi_m\rangle$ could be represented via outer products of
one-dimensional Fock states $|k\rangle$
 \begin{equation}
 \label{eq-dec-Fock}
 |\psi_m\rangle = \sum_{k_1,k_2,\ldots,k_n}
    A_{k_1,k_2,\ldots,k_n}^{(m)} \,
    |k_1\rangle \, |k_2\rangle \ldots |k_n\rangle,
 \end{equation}
where $k_i$, $i = 1, 2, \ldots, n$ corresponds to $i$th
one-dimensional subspace.

The right-hand side of the uncertainty relation (\ref{eq-triv-nd})
is calculated using the method of
papers~\cite{Bastiaans1983a,DodonovManko1989}. Core idea of
calculation is the introduction of auxiliary observable
 \begin{equation}
 \label{eq-energ}
 E(\vartheta) = \frac{1}{2} \left[(\Delta P)^2/\vartheta +
     \vartheta (\Delta X)^2\right],
 \end{equation}
which could be regarded as energy of some oscillator with unit
frequency and mass $\vartheta$. The minimum of $E(\vartheta)$ with
respect to $\vartheta$ (for $\vartheta = \Delta X/\Delta P$) is
exactly left-hand side of uncertainty relation
 \[
 \min_\vartheta E(\vartheta) = \Delta X \, \Delta P.
 \]
As far as Fock states are eigenstates of the harmonic oscillator
with eigenvalue $2k + 1$ (in the one-dimensional case), then
substitution of (\ref{eq-spectral}) and (\ref{eq-dec-Fock}) into
(\ref{eq-energ}) leads to
 \begin{equation}
 \label{eq-general}
 \Delta X\,\Delta P \geqslant \frac{\hbar}{2} \: \frac{1}{n}
 \sum\limits_m \rho_m \sum_{k_1,k_2,\ldots,k_n}
 \left[2(k_1 + k_2 + \ldots k_n) + n\right]\;
 \left|A_{k_1, k_2, \ldots, k_n}^{(m)}\right|^2.
 \end{equation}

Now, due to isomorphism between the set of all positive integers
and the set of combinations of $n$ positive integers,
it is possible to consider the coefficients $A_{k_1, k_2, \ldots,
k_n}^{(m)}$ as elements of some unitary matrix
$\{\tilde{A}_{\mathbf{k} \, m}\}$. Then, (in)equality
(\ref{eq-general}) can be treated with lemma\footnote{See also
\cite{Dodonov2002}; to be self-contained, the lemma is reproduced,
together with necessary changes for multidimensional case, in 
\ref{lemma} of this article.} from \cite{Bastiaans1984} to give
 \begin{equation}
 \label{eq-u3-3a}
 \Delta X\,\Delta P \geqslant \frac{\hbar}{2} \: \frac{1}{n}
 \sum_{k_1,k_2,\ldots,k_n} \left[2(k_1 + k_2 + \ldots k_n) + n\right]\;
 \rho_{m(k_1,k_2,\ldots,k_n)},
 \end{equation}
where eigenvalues of the density matrix are ordered in a
non-increasing sequence.

Dependence of the expression $2(k_1 + k_2 + \ldots + k_n) + n$ on
indices $k_1,k_2,\ldots,k_n$ is degenerate: this expression takes
the same values for several combinations of indices. Therefore
it is possible to rewrite the (in)equality (\ref{eq-u3-3a}) as
 \begin{equation}
 \label{eq-u3-3}
 \Delta X\,\Delta P \geqslant \frac{\hbar}{2} \: \frac{1}{n}
 \sum_k (2k + n) \sum_{m = 0}^{g^{(n)}_m - 1} \rho_{m(k)}
 \end{equation}
where the values $g_m^{(n)}$ (degeneration multiplicity) are defined
by the formula (\ref{eq-degener}), and the eigenvalues of the density
matrix are collected in groups of $g_m^{(n)}$ terms. Expression
(\ref{eq-u3-3}) is the main result of the paper, and it is the most
general form of the uncertainty relation for mixed states (partially
coherent fields) in a multidimensional space. This inequality
relates a minimal uncertainty volume of a state to the spectrum of
the density operator corresponding to this state.

 \section{Multidimensional purity-bounded relations}
 \label{Gen-bound}

Using the method from papers \cite{Bastiaans1984,Bastiaans1986a},
it is possible to find a dependence of uncertainty relation limit
on some characteristics of purity of quantum system. Usually,
a family of ``generalized purities'' (Shatten $p$-norms
or ``generalized entropies'' \cite{Adesso_etal2004}),
is used, which is defined as
 \begin{equation}
 \label{eq-mu_r}
 \mu^{(r)} = \left[ \Tr \left(\hat\rho^{\,r/(r-1)}\right) \right]^{r-1},
 \end{equation}
where $r$ is an arbitrary (not necessary integer) real number with
$r>1$. Important special cases of $\mu^{(r)}$ include $\mu^{(2)}=\mu$
(``usual'' purity, see above), ``superpurity''
 \begin{equation}
 \label{eq-mu_1}
 \mu^{(1)} = \lim_{r \to 1} \mu^{(r)},
 \end{equation}
when only the largest eigenvalue of the density matrix is taken
into account, and also ``entropy-based'' purity degree
 \begin{equation}
 \label{eq-mu_ent}
 \mu_S = \exp(-S),
 \quad
 S=-\Tr(\hat\rho \ln \hat\rho),
 \end{equation}
which is defined in terms of Shannon-–von Neumann entropy $S$, and so
leads to ``entropy-bounded'' uncertainty relation. As it is shown in
\cite{Bastiaans1986b}, $\mu_S$ can be treated as a limiting case
of definition (\ref{eq-mu_r}) for $r \to \infty$: $\mu^{(\infty)}
= \mu_S$.

``Superpurity'' and entropy-bounded uncertainty relations play a
special role for one dimensional case: owing to continuous
non-increasing dependence of $\mu^{(r)}$ on $r$ for $r \ge 1$
\cite{Bastiaans1986b}, they are limiting cases of family of
characteristics (\ref{eq-mu_r}). It is also possible to show, that
the non-increasing dependence of $\mu^{(r)}$ on $r$ remains valid
in a general multidimensional case, see details in 
\ref{Nonincreasing}.

As it can be easily shown by Lagrange method, at the minimum of
uncertainty relation, the definition (\ref{eq-mu_r}) reduces to
 \begin{equation}
 \label{eq-u3-7}
 \mu = \left[
     \sum_k g_k^{(n)} \; (\xi_k/g_k^{(n)})^{r/(r-1)}
 \right]^{r-1},
 \end{equation}
where
 \[
 \xi_k = \sum_{m = 0}^{g_s^{(n)} - 1} \rho_{m(k)}
 \]
and
 \begin{equation}
 \label{eq-xi-norm}
 \sum_k \xi_k=1.
 \end{equation}
Then the (in)equality (\ref{eq-u3-3}) may be rewritten to
 \begin{equation}
 \label{eq-u3-3b}
 \Delta X \, \Delta P \geqslant \frac{\hbar}{2} \: \frac{1}{n}
 \sum_k (2k + n) \, \xi_k
 \end{equation}
and, in order to obtain the relation of the type (\ref{eq-asymp})
for given structure of the density matrix, (i.~e. eigenvalues
$\rho_m$) it is necessary to find a minimum with respect to
variables $\xi_s$.


As the degeneration multiplicity (\ref{eq-degener}) has a
rather complex form, the task of detailed study of uncertainty
relation minimum has, in general, no analytical solution (the same as
in one-dimensional case \cite{Dodonov2002}). Besides interpolated
and asymptotic inequalities, which will be studied later in the
article, it is possible to obtain analytical solution for the case of
entropy-bounded relations.

Using Lagrange method, it is easy to show, that a minimum of
uncertainty product $\Delta X \Delta P$ (\ref{eq-u3-3b}) for given
entropy $S$ is attained if the coefficients $\xi_m$ are given by
 \begin{equation}
 \label{eq-ent-xi}
 \xi_m = A \: g_m^{(n)} \exp(-\beta m),
 \end{equation}
where $A$ is a normalization constant and parameter $\beta$
depends on the entropy. Taking into account structure of the
density operator at the minimum of general uncertainty relation
(\ref{eq-u3-3}) [compare representation (\ref{eq-dec-Fock})], in the
case of entropy-bounded relation the density matrix is taking the
form
 \begin{equation}
 \hat\rho^{(n)}_S = \underbrace{\hat\rho^{(1)}_S \;
    \hat\rho^{(1)}_S \ldots \hat\rho^{(1)}_S}_{\mbox{$n$ times}}.
 \end{equation}
Here
 \begin{equation}
 \label{eq-thermal}
 \hat\rho^{(1)}_S = \sum_{k=0}^\infty \rme^{-\beta k} \,
 |k \rangle \langle k|
 \end{equation}
is a density matrix corresponding to minimum of one dimensional
entropy-bounded relation. Parameter $\beta$ can be found from
solution of transcendental equation
 \begin{equation}
 \label{eq-u3-9}
 S = \beta \, n \, \frac{\exp(\beta)}{1 - \exp(\beta)} -
     n \, \ln\left(1 - \exp(- \beta)\right),
 \end{equation}
which is in accordance with appropriate equation for
one-dimensional case \cite{DodonovManko1989}.

Uncertainty relation then could be written as
 \begin{equation}
 \label{eq-u3-8}
 (\Delta X\;\Delta P)^n \geqslant \beta(S) \,
 \left(\frac{\hbar}{2}\right)^n,
 \end{equation}
or, for highly-mixed states with $S \gg 1$
 \begin{equation}
 \label{eq-u3-8a}
 (\Delta X\;\Delta P)^n \geqslant \exp(S)
 \left(\frac{2}{\rme}\right)^n\,
 \left(\frac{\hbar}{2}\right)^n.
 \end{equation}
Obtained structure of eigenstate decomposition is factorized
on solutions of one dimensional problem (see
\cite{DodonovManko1989,Bastiaans1986a}), which leads to thermal
state (\ref{eq-thermal}). In other words, the entropy-bounded
uncertainty relation of position-momentum type has no additional
effects for multidimensional cases. 


In order to study general low-purity case, it is possible to
utilize the approach from Bastiaans' paper \cite{Bastiaans1984},
which is based on generalization of the H\"older inequality.
The mathematical details are presented in 
\ref{Hoelder}.

Let's define an ``uncertainty function'' $C(\mu^{(r)},n)$ (see
also (\ref{eq-asymp})) as
 \begin{equation}
 \label{eq-C-gen}
 C(\mu^{(r)},n) = \mu^{(r)} \left( \frac{1}{n}
    \left\{2\,M + n - 2 \left[\mu^{(r)} \;
       B(M,n,r)\right]^{1/r} \right\} \right)^n,
 \end{equation}
where
 \begin{equation}
 \label{eq-Br}
 B(M,n,r)=\sum_{0\leqslant m \leqslant M}
          {\frac{(m+n-1)!}{(n-1)!\, m!} (M-n)^r}
 \end{equation}
and additional real and positive minimization parameter $M$ is
introduced.

\begin{figure}[t!]
 \centerline{\scalebox{.9}{\includegraphics{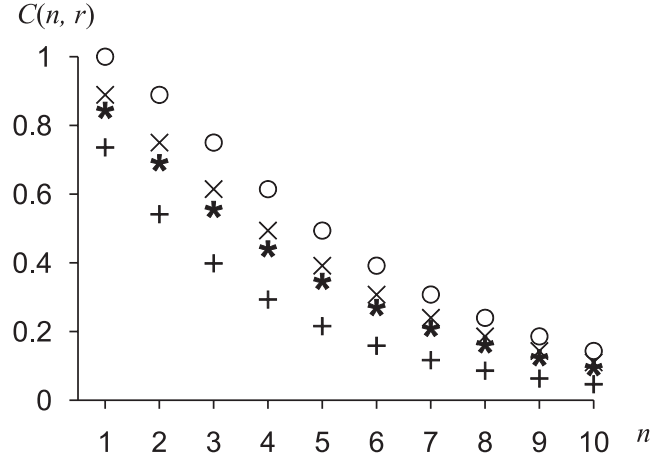}}}
 \caption{Uncertainty relation minimum for $\mu^{(r)} \ll 1$,
$n=1,\,\ldots\,6$ and different variants of the degree of purity:
$\circ$'s --- $r \to 1$, $\times$'s --- $r = 2$, $\ast$'s --- $r =
3$, $+$'s --- $r \to \infty$ (entropy-bounded relation).}
 \label{fig1}
\end{figure}

\begin{figure}[ht!]
 \centerline{\scalebox{.9}{\includegraphics{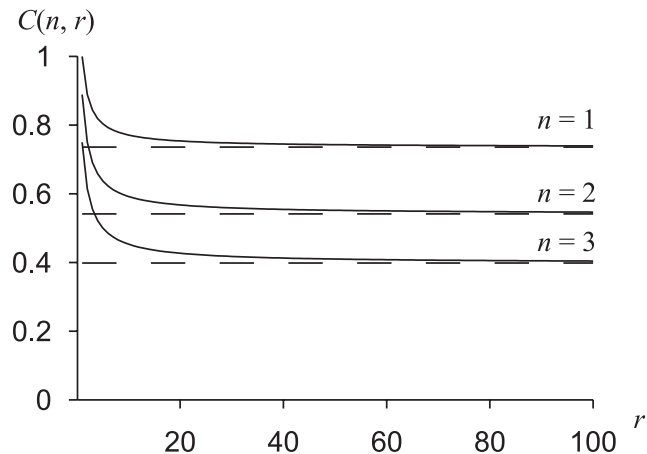}}}
 \caption{Uncertainty relation minimum for $\mu^{(r)} \ll 1$,
$n=1,\,2,\,3$ and different variants of the degree of purity
(solid lines, $r = 1, \ldots, 100$). Dashed lines denote
uncertainty minimum for entropy-bounded relation with
$r \to \infty$ (for $n=1,\,2,\,3$, rspectively).}
 \label{fig2}
\end{figure}

1-d problem has known asymptotical solution \cite{Bastiaans1984}
 \[
 C(\mu^{(r)},1) = 2 \, [r/(r+1)]^r,
 \qquad
 \mu^{(r)} \ll 1.
 \]
In the same way, highly mixed states could be treated analytically
for arbitrary multidimensional case: as far as limit of small
$\mu^{(r)}$ requires $M$ to be sufficiently
large~\cite{KarelinLazaruk2000,Dodonov2002}, it is possible to
replace summation in formula (\ref{eq-Br}) by integration together
with approximation of degeneration multiplicity by $m^{n-1}/(n-1)!$
with
 \begin{equation}
 \label{eq-Br1}
 B(M,n,r) \approx \frac{1}{(n-1)!} \int_0^M m^{n-1} (M-m)^r \: \rmd m.
 \end{equation}
The last relation could be calculated analytically,
 \begin{equation}
 \label{eq-Br2}
 B(M,n,r) \approx M^{n+r+1} \left(\prod_{k=1}^{n+1} (r+k)\right)^{-1}
 \end{equation}
(see 
\ref{Br1-int} for details). Further minimization of relation
(\ref{eq-C-gen}) with respect to $M$, after some tedious but quite
elementary algebra gives an asymptotic variant of general purity
bounded uncertainty relation
 \begin{equation}
 \label{eq-pur-bound}
 (\Delta X \, \Delta P)^n \geqslant
   \left(\frac{\hbar}{2}\right)^n \: \frac{C(n,r)}{\mu^{(r)}},
 \qquad
 C(n,r) = \frac{2^n \, r^r}{(n+r)^{n+r}} \:
   \prod_{k=1}^n (r+k),
 \end{equation}
which describes the whole range of $n$ and $r$ for $\mu^{(r)} \ll
1$. Resulting dependencies of $C(n,t)$ on $r$ for $n = 2,3$
together with one-dimensional case are presented in
figure~\ref{fig1} and figure~\ref{fig2}. It is seen, that the
obtained expressions correctly describe whole range of parameter $r$,
demonstrating decrease of uncertainty minimum with increase of $r$
and leading to entropy-bounded relations at $r \to \infty$.

\section{Concluding remarks}

To summarize, it is worth to note that two main forms of
uncertainty principle (of position-momentum type) for mixed states
in multidimensional space are obtained in the present paper. The
first one (\ref{eq-u3-3}) relates minimal uncertainty product with
eigenspectrum of the density matrix and the second
(\ref{eq-pur-bound}) is a general (asymptotic) purity-bounded
uncertainty relation. In both cases, minimum of uncertainty
product is obtained when eigenstates of the density operator are
Fock states. In the case of purity-bounded relation, eigenspectrum
of the density operator is defined by
 \begin{equation}
 \xi_m \propto g_m^{(n)} (M-m)^{r-1},
 \quad
 0 \leqslant m \leqslant M
 \end{equation}
see 
\ref{Hoelder}. In other words, spectral representation of the
density matrix is a finite sum of Fock states. Such state is
definitely non-classical, see discussion in the Dodonov's
paper~\cite{Dodonov2002}. Upon transition to $r \to \infty$
(entropy-bounded relations), the minimum-uncertainty state becomes
thermal (\ref{eq-thermal}), i.~e. classical. More detailed study
of minimum-uncertainty states structure will be subject of another
publication.

It is also necessary to note, that the results of the paper is
applicable for analysis and characterization of entangled quantum
states. Indeed, the spectral decomposition of density matrix is
closely connected to the Schmidt decomposition of non-separable
states, see, e.~g.~\cite{Bruss2002}. Approach to uncertainty
principle for entangled states can be based on mathematically
analogous case of uncertainty (reciprocity) relations for pulsed
partially coherent classical beam~\cite{KarelinLazaruk2002}.

\ack

 Author want to thank Dr.~Dmitry Mogilevtsev and Dr.~Vya\-che\-slav
 Shatokhin (B.~I.~Stepanov Institute of Physics) for stimulating
 discussions and critical remarks.

\appendix
 \def\thesection{Appendix~\Alph{section}}

 \section{}

 \label{lemma}

The lemma is reproduced here mainly for completeness of material,
and in order to make the above analysis clearer. Initially it was
presented in Appendix~A of paper \cite{Bastiaans1984}, it also
could be found in \cite{Dodonov2002}. According to
\cite{Bastiaans1984}, idea of this proof was initially proposed by
M.~L.~J.~Hautus.

Let the sequence of numbers $b_m$ be defined by
 \begin{equation}
 b_m = \sum_{k = 0}^{\infty} |a_{mk}|^2 \gamma_k,
 \end{equation}
where $\gamma_0 \leqslant \gamma_1 \leqslant \ldots \leqslant
\gamma_k \leqslant \ldots$ and coefficients $a_{mk}$ satisfy the
orthonormality condition
 \begin{equation}
 \sum_{k = 0}^{\infty} a_{mk} \, a_{lk}^{\ast} = \delta_{ml},
 \quad m,l = 0,1,\ldots
 \end{equation}
One may consider the numbers $b_m$ for $m = 0,1,\ldots,M$ as the
diagonal entries of an $(M+1)$-square Hermitian matrix $H =
\|h_{ij}\|$ with
 \begin{equation}
 h_{ij} = \sum_{k = 0}^{\infty} a_{ik} a_{jk}^{\ast} \gamma_k,
 \quad i,j = 0,1,\ldots,M.
 \end{equation}
Let the eigenvalues $\beta$ of $H$ be ordered according to
 \begin{equation}
 \beta_0 \leqslant \beta_1 \leqslant \ldots
 \leqslant \beta_k \leqslant \ldots \leqslant \beta_M.
 \end{equation}
From Cauchy's inequalities for eigenvalues of a submatrix of a
Hermitian matrix \cite{MarcusMinc1972}, 
we can conclude, that $\beta_m \geqslant \gamma_m$ $(m =
0, 1, \ldots, M)$ and hence
 \begin{equation}
 \sum_{m=0}^M b_m = \sum_{m=0}^M h_{mm} = \sum_{m=0}^M \beta_m
 \geqslant \sum_{m=0}^M \gamma_m.
 \end{equation}
Furthermore, with the numbers $\lambda_m$ (or $\rho_m$, in this
article) satisfying the property $\lambda_0 \geqslant \lambda_1
\geqslant \ldots \geqslant \lambda_m \geqslant \ldots$, we can
formulate the chain of relations
 \begin{equation}
 \eqalign{
 \sum_{m=0}^M \lambda_m b_m
    & = \lambda_0 b_0 + \sum_{m=1}^M \lambda_m b_m 
    = \lambda_0 b_0 + \sum_{m=1}^M \lambda_m
         \left[ \sum_{l=0}^m b_l - \sum_{l=0}^{m-1} b_l \right] \\
    & = \lambda_0 b_0 + \sum_{m=1}^M \lambda_m \sum_{l=0}^m b_l -
         \sum_{m=0}^{M-1} \lambda_{m+1} \sum_{l=0}^m b_l \\
    & = \sum_{m=0}^{M-1} \lambda_m \sum_{l=0}^m b_l + \lambda_M \sum_{l=0}^M b_l -
         \sum_{m=0}^{M-1} \lambda_{m+1} \sum_{l=0}^m b_l \\
    & = \lambda_M \sum_{l=0}^M b_l +
         \sum_{m=0}^{M-1} (\lambda_m - \lambda_{m+1}) \sum_{l=0}^m b_l \\
    & \geqslant \lambda_M \sum_{l=0}^M \gamma_l +
         \sum_{m=0}^{M-1} (\lambda_m - \lambda_{m+1}) \sum_{l=0}^m \gamma_l
         = \sum_{m=0}^M \lambda_m \gamma_m.
 }
 \end{equation}

On choosing $\gamma_n = 2 n + 1$ and taking the limit $M \to
\infty$, we arrive at the inequality
 \[
 \sum_{m=0}^{\infty} \lambda_m
      \sum_{n=0}^{\infty} |a_{mn}|^2 (2n + 1)
 \geqslant
 \sum_{m=0}^{\infty} \lambda_m (2m + 1),
 \]
which becomes an equality if $|a_{mn}| = \delta_{mn}$.

In order to modify this proof to miltidimensional case, it is
necessary to choose $\gamma_m$ as $\gamma_\vcm = 2(m_1 + \ldots +
m_s) + n$ (here $\vcm=(m_1, \ldots, m_n)$ is a ``vectorial'' summation
index), and then to take into account degeneracy of coefficients
$\gamma_\vcm$.

 \section{}
 \label{Nonincreasing}

By analogy with Bastiaans' paper \cite{Bastiaans1986b} for any
$r$, $q$, with $1 < r < q$, holds
 \begin{equation}
 \label{eq-r-prop}
 \fl \eqalign{
 \mu^{(r)} & = \left[\sum_m g_m^{(n)} \theta_m^{r/(r-1)}\right]^{r-1}
 = \left[ \sum_m g_m^{(n)}
 \left(\theta_m^{q/(q-1)}\right)^{(q-1)/(r-1)}
 \left(\theta_m\right)^{(r-q)/(r-1)}
   \right]^{r-1} \\
 & \leqslant
 \left[ \left( \sum_m g_m^{(n)} \theta_m^{q/(q-1)}
        \right)^{(q-1)/(r-1)} \;
 \left( \sum_m g_m^{(n)} \theta_m \right)^{(r-q)/(r-1)}
 \right]^{r-1} = \mu^{(q)}.
 }
 \end{equation}
Here $\theta_m = \xi_m/g_m^{(n)}$, and the H\"older inequality for
weighted sum \cite{MathEncycl} is used, see formula
(\ref{eq-Hoeld-3}). Therefore, for a family of purities (\ref{eq-mu_r}),
it is possible to conclude, that ``superpurity'' and entropy-based
purity lead to limiting cases of all multidimensional uncertainty
relations.

 \section{}
 \label{Hoelder}

Starting from the equation (\ref{eq-xi-norm}), with $\xi_m$ a sequence
of nonnegative numbers and $M$ an arbitrary real nonnegative
constant, we write
 \begin{equation}
 \label{eq-Hoeld-1}
 \sum_{m=0}^{\infty} \xi_m = \frac{1}{2M + n}
    \left[2 \sum_{m=0}^{\infty} \xi_m (M - m) +
    \sum_{m=0}^{\infty} \xi_m (2m + n)
    \right] = 1.
 \end{equation}

The following (in)equalities hold:
 \begin{equation}
 \label{eq-Hoeld-2}
 \fl \sum_{m=0}^{\infty} \xi_m (M - m) \leqslant
 \sum\limits_{0\leqslant m \leqslant M} \xi_m (M - m),
 \end{equation}
 \begin{equation}
 \label{eq-Hoeld-3}
 \fl \eqalign{
 \sum\limits_{0\leqslant m \leqslant M} \xi_m (M-m)
 & = \sum\limits_{0\leqslant m \leqslant M} g_m^{(n)} (M-m)
    \xi_m/g_m^{(n)} \\
 & \leqslant \left[ \sum\limits_{0\leqslant m \leqslant M}
    g_m^{(n)} (M-m)^r \right]^{1/r} \;
 \left[ \sum\limits_{0\leqslant m \leqslant M} g_m^{(n)}
    \left(\xi_m/g_m^{(n)}\right)^p
 \right]^{1/p},
 }
 \end{equation}
 \begin{equation}
 \label{eq-Hoeld-4}
 \fl \sum\limits_{0\leqslant m \leqslant M} g_m^{(n)}
    \left(\xi_m/g_m^{(n)}\right)^p
 \leqslant
 \sum_{m=0}^{\infty} g_m^{(n)} \left(\xi_m/g_m^{(n)}\right)^p
 \end{equation}
with two real parameters $p, r \geqslant 1$, $1/p + 1/r = 1$. The
equality sign in relations (\ref{eq-Hoeld-2}) and
(\ref{eq-Hoeld-4}) holds if $\xi_m = $ for $m > M$. Relation
(\ref{eq-Hoeld-3}) changes to equality, when
$\xi_m \propto g_m^{(n)} (M-m)^{r-1}$ 
in the interval $0\leqslant m \leqslant M$. (In)equality
(\ref{eq-Hoeld-3}) is a general form of the H\"older inequality
for the weighted sum \cite{MathEncycl}, see also
\cite{Mitrinovic1970}.

Combining the (in)equalities (\ref{eq-Hoeld-1}) --
(\ref{eq-Hoeld-4}) gives a relation
 \begin{equation}
 \label{eq-Hoeld-5}
 \frac{1}{2M + n} \left[2 B(M,n,r)^{1/r} \mu_q +
     \sum_{m=0}^{\infty} \xi_m (2m + n) \right] \geqslant 1,
 \end{equation}
where
 \begin{equation}
 \label{eq-Hoeld-6}
 B(M,n,r) = \sum\limits_{0\leqslant m\leqslant M}
    g_m^{(n)} (M - m)^r,
 \end{equation}
 \begin{equation}
 \label{eq-Hoeld-mu}
 \mu_p = \left[ \sum_{m=0}^{\infty}
    g_m^{(n)} \left(\xi_m/g_m^{(n)}\right)^p
 \right]^{1/rp}.
 \end{equation}

From the condition $1/p + 1/r = 1$ it follows that $p = r/(r-1)$ and
then the (in)equality (\ref{eq-C-gen}) results.

 \section{}
 \label{Br1-int}

In order to find an integral in approximation (\ref{eq-Br1}), we
start from introduction of a new variable $x = M-m$
 \begin{equation}
 B(M,n,r) \approx \frac{1}{(n-1)!} \int_0^M \rmd x \: (M-x)^{n-1} x^r.
 \end{equation}
Application of binomial formula to $(M-x)^{n-1}$ and interchanging the
order of integration and summation leads to
 \begin{equation}
 B(M,n,r) \approx
 M^{n+r+1} \sum_{k=0}^n \frac{(-1)^k}{k! \, (n-k)! \, (k+r+1)}
 \end{equation}
The last  sum can be calculated by use of formula (5.41) from
Graham, Knuth and Patashnik book~\cite{Graham_etal1994}, leading
at last to~(\ref{eq-Br2}).


\end{document}